\documentstyle[12pt,a4]{article}
\vspace*{-2.5cm}
\newcommand{\half}{\mbox{${\textstyle \frac{1}{2}}$}}           
\def\fmn#1#2{\mbox{${\textstyle \frac{#1}{#2}}$}}
\begin{document}
\vspace{1cm}

\begin{center}
{\LARGE\bf The \mbox{\boldmath\LARGE $dd\rightarrow \,^4\!H\!e\,\eta$}
reaction near threshold}\\[8ex]
G\"oran F\"aldt\footnote{Electronic address faldt@tsl.uu.se}\\[1ex]
Division of Nuclear Physics, Uppsala University, Box 535,
S-751 21 Uppsala, Sweden\\[4ex]
Colin Wilkin\footnote{Electronic address cw@hep.ucl.ac.uk}\\[1ex]
University College London, London, WC1E 6BT, UK\\[8ex]
\end{center}

\begin{abstract}
The cross section for the reaction $dd\rightarrow \,^4\!H\!e\,\eta$ close to
threshold is estimated in a two-step model, similar to that which
successfully describes near-threshold $pd\rightarrow \,^3\!H\!e\,\eta$
production. A $\pi$-meson, created in a
$dp\rightarrow \,^3\!H\!e\,\pi^o$ process on one nucleon in the target
deuteron, is converted into an $\eta$-meson by a secondary
$\pi^o n \to n\,\eta$ reaction on the other nucleon. The kinematics are such
that both processes are nearly physical so that only small Fermi momenta are
required. The predictions of the model are in good agreement with recent
Saclay experimental data.

An analogous model, when applied to the near-threshold
$n\,^3\!H\!e\rightarrow \,^4\!H\!e\,\eta$ reaction, predicts a cross section of
a rather similar size, but experimental difficulties render this a doubtful
means of studying the $\eta\, ^4\!H\!e$ interaction.
\end{abstract}
\vspace{2cm}
\centerline{\today}

\newpage
\section{Introduction}

The strong energy variation observed in the $pd\rightarrow \,^3\!H\!e\,\eta$
amplitude near threshold \cite{Berger,Garcon} lead to the speculation of
the existence of quasi-bound $\eta$-nucleus states \cite{CW1} for much
lighter nuclei than originally suggested \cite{Liu}. This incited groups
working at the SPESIV \cite{Roudot} and SPESIII \cite{Amina} spectrometers at
Saturne to investigate the next heavier nucleus through measurements of the
$dd\rightarrow \,^4\!H\!e\,\eta$ reaction. First results indicate a
$^4\!H\!e\,\eta$ threshold production cross section which is about a factor
of 40 less than for $^3\!H\!e\,\eta$, combined with a weaker energy
dependence.

The near-threshold energy variation is dominated by final-state-interaction
effects, which reflect the $\eta$-nucleus dynamics, and these should be
largely independent of the particular entrance channel. In contrast, the
prime aim of the present paper is to try to understand the factor of 40
between the two production rates, and this leads us rather to studying the
reaction mechanism itself.

Near-threshold production of a meson as massive as the $\eta$ necessarily
involves large momentum transfers and models which rely
on spectator nucleons yield much too low cross sections \cite{LagLec,GW2}.
In a previous paper \cite{FW1}, hereinafter referred to as I, we
presented and evaluated a two-step model for the
$pd\rightarrow \,^3\!H\!e\,\eta$ reaction, based on ideas put forward by
Germond \cite{JF} and Kilian and Nann \cite{Kil}. The basic assumption is
that a virtual pion beam is created by the interaction of beam protons
with one of the nucleons in the target and it is this pion beam which
produces the $\eta$, through an interaction with the second nucleon in the
target. A similar model has been independently studied in reference
\cite{Kond}.

The importance of this two-step mechanism is enhanced by what has been
called the {\it kinematic miracle} by Kilian and Nann\cite{Kil}. They noticed
that the kinematics are such that both sub-processes can proceed at very
small Fermi momenta, with the constituent particles being almost on their
energy
shells. Any quantitative evaluation of such a model should not therefore be
plagued by off-shell uncertainties and reliance on tails of Fermi momentum
distributions. Taking advantage of the nearness to the physical domain,  we
expanded the intermediate state propagator to first order in the Fermi momenta
to obtain a transparent physical estimate of the
$pd\rightarrow \,^3\!H\!e\,\eta$ cross section in terms of those for
$pp\to d\pi^+$ and $\pi^+n\to\eta p$ times simply evaluated form factors.

Close to threshold the only significant energy dependence comes from the
final state interactions. After taking these crudely into account, the model
predicts cross sections which are too low by about a factor\footnote{Due to a
numerical approximation, the original factor of $2.5$ in I
should be corrected to $2.8$.} of $2.8$, which
we attribute to the neglect of contributions of intermediate $NN$ singlet
states
($d^*$). This is consistent with the observation that, with the same
normalisation factor, the gross features of the threshold production rates for
$\omega$, $\eta'$ and $\phi$ mesons in the $pd\rightarrow \,^3\!H\!e\,X$
reaction are equally well explained by the model \cite{FW2}.

In the present paper, we extend the model by considering the
$dd\rightarrow \,^4\!H\!e\,\eta$ reaction to be the result of a
$dp\rightarrow \,^3\!H\!e\,\pi^o$ process on one target nucleon, followed by
$\pi^o n \to n\,\eta$ reaction on the other. The predictions of this model
in the forward direction near threshold,
including final-state-interaction effects, are even closer to experimental
data than for the lighter isotope, giving extra credence to the virtual pion
beam concept.

The formalism is completely analogous to that in I, so that the
development given here can be brief. After presenting the generic form of
the two-step diagram in \S2, the amplitude is written as an integral over
the $^4\!H\!e$ and deuteron Fermi momenta. By linearising the propagator in
\S3, the six-dimensional integral is reduced to a single one and an explicit
prediction obtained for the $dd\rightarrow \,^4\!H\!e\,\eta$ cross section.
The effective wave technique used in I for handling the $\eta$-nucleus
final state interactions is generalised in \S4 to the case of $^4\!H\!e\,\eta$.

An alternative way of obtaining the same exit channel, in order to study the
$^4\!H\!e\,\eta$ final state interaction, is through the measurement of the
$n\,^3\!H\!e\rightarrow \,^4\!H\!e\,\eta$ reaction. This can be estimated
in a similar two-step mechanism, and the method for doing this is sketched
in \S5, where an explicit formula is presented for this cross section
in terms of the same $pd\rightarrow \,^3\!H\!e\,\pi^o$ and
$\pi^o n \to n\,\eta$ cross sections that arise in the $dd$ case.

Our results and conclusions are presented in \S 6, where it is seen that the
overall magnitude of our $dd\rightarrow \,^4\!H\!e\,\eta$ predictions agrees
even better with the experimental data than we found for
$pd\rightarrow \,^3\!H\!e\,\eta$, possibly because of the absence of any
unquantifiable $d^*$ contributions in the former case. On the other hand our
treatment of the final state interactions leads to too sharp a threshold
behaviour.

Our $n\,^3\!H\!e\rightarrow \,^4\!H\!e\,\eta$ cross section estimate is
of the same order of magnitude as for the deuteron-initiated
reaction. Taking also into account the difficulty of working with a
neutron beam or a tritium target, it suggests that this reaction is not
competitive with $dd\rightarrow \,^4\!H\!e\,\eta$ for producing low energy
$\eta$-mesons.

\newpage
\section{Nuclear matrix element}

In terms of nuclear matrix elements $\cal M$ with the same normalisation
as in  I, the c.m.~differential cross section for the unpolarised
$dd\rightarrow \,^4\!H\!e\,\eta$ reaction is written
\begin{equation}
\frac{d\sigma}{d\Omega}=\frac{1}{16(2\pi)^{2}s}\frac{p_{\eta}}{p_d}\,
\frac{1}{9} \sum_{\mbox{\scriptsize spins}}|{\cal M}|^{2}\ ,
\label{total_DCS}
\end{equation}
with $\sqrt{s}$ being the total c.m.~energy of the
${\eta}\,^4\!H\!e$-system and $p_{\eta}$ and $p_d$ the momenta of the
$\eta$ and deuteron in this system.

There are strong restrictions on the amplitude structure at
threshold for the $\eta\alpha$ system has there $J^p = 0^-$, which necessitates
the deuterons to be in a $L=1$, $S=1$ state. The unique amplitude may be
written in invariant form as
\begin{equation}
{\cal M}=\frac{1}{(2\pi)^3}\,\sqrt{4m_{\alpha}\,m_d} \,
\left[iF\,(\vec{\epsilon}_d\,'\times\vec{\epsilon}_d)\cdot\hat{p}_d\right]
\: , \label{uppdelM}
\end{equation}
where the $\vec{\epsilon}_d$ and $\vec{\epsilon}_d\,'$ are the standard
orthonormal sets of deuteron polarisation vectors.

Though actually symmetric under the exchange of the two deuterons, at
threshold we did not need to impose this in order to derive the spin structure
of $\cal M$ in eq.(\ref{uppdelM}). However in the forward direction, even away
from threshold, it is in fact the \underline{only} amplitude with the correct
Bose symmetry. This means that in the forward direction, only deuterons with
helicity $m=\pm 1$ can initiate the reaction and so there is a unique deuteron
tensor analysing power, $t_{20} = +1/\sqrt{2}$. This result is used in the
analysis of the SPESIII experiment \cite{Amina}, which employed a tensor
polarised deuteron beam.

In terms of the amplitude $F$, the differential cross section of
eq.(\ref{total_DCS}) becomes
\begin{equation}
\left[\frac{d\sigma}{d\Omega}(dd\rightarrow \,^4\!H\!e\,\eta)\right]_{cm}
= \frac{m_{\alpha}\,m_d}{18\,(2\pi)^{8}s}\, \frac{p_{\eta}}{p_d}
\,{\mid}F{\mid}^{2}\:.\label{dd_cs}
\end{equation}

In analogy with the model proposed in I, one of the two-step contributions
to the $dd\rightarrow \,^4\!H\!e\,\eta$ is where a secondary $\pi^0$ beam is
created through the $dp\to \,^3\!H\!e\,\pi^0$ reaction on the proton in the
target deuteron. The $\pi^0$ is subsequently transformed into an
$\eta$ by a $\pi^0 n\to\eta n$ interaction on the target neutron.
This model is expressed in diagrammatic terms in fig.~1.

The nuclear matrix element for the reaction can be written as
\begin{equation}
{\cal M}=N_T\,\sqrt{4m_{\alpha}m_d}\,\int\,\frac{m_h}{E_h}\,d^{\,3}k\,
\int\,\frac{m_n}{E_n}\,d^{\,3}q\ \frac{1}{(2\pi)^3\,2E_{\pi}}
  \:\frac{i}{E_{0}-E_{int}+i\epsilon}\ {\cal M}_{K}\: , \label{basM}
\end{equation}
where $E_{0}=E_{\eta}+E_{\alpha}$ is the total energy of the initial
(and final) state. $E_n$ and $E_h$ are the energies of the neutron and
helion in the intermediate state, where the total energy is $E_{int}$.

${\cal M}_{K}$ is the product of individual matrix elements;
\begin{equation}
{\cal M}_{\rm K}=\sum_{\mbox{\scriptsize int}}\,
            {\cal M}(n'h_c\to {\alpha}_{f}\,)
            {\cal M}(\pi^{0}n\to n'\eta_{f})\,
            {\cal M}(d_{i}\to p_{c}n)\,
            {\cal M}(d_{i}'p_{c}\to\pi^{0}h_c)\: ,
\label{nuclMk}
\end{equation}
where the sum runs over the spin degrees of freedom of the intermediate
particles. An index $c$ signifies that the particle is represented by
its charge conjugate spinor.

In   addition to  the  process   described by     eq.(\ref{nuclMk}),  there is
a
contribution of  similar size coming  from the graph where  protons and
neutrons
are interchanged,  and two more  contributions from charged  intermediate
pions.
The latter are larger by a factor of  two and so the amplitude of fig.~1 must
be
increased  by an  isospin  factor  of six.  However, there are in addition an
identical set of graphs where beam and target deuterons are interchanged and,
in
the forward  direction, these  ensure the  correct $m=\pm 1$  helicity
selection
rule even away from threshold. Thus the  $m=\pm 1$ amplitudes of
eq.(\ref{basM})
must be  multiplied by a  total  combinatorial  factor of  $N_T=12$, whereas
the
$m=0$ amplitudes vanish.

We now study separately the four matrix elements of eq.(\ref{nuclMk})
\subsection{The deuteron wave function}

The momentum space wave function for an initial state  deuteron at rest is
\begin{eqnarray}
{\cal M}(d_i\to p_{c}n)=
\mbox{\large $\eta$}_{n}^{\dagger}\frac{1}{\sqrt{2}}
\left[ -\vec{\sigma}\cdot\vec{\epsilon}_{d}\,\varphi_{S}(q)+
\frac{1}{\sqrt{2}}
\left(3\vec{\sigma}\cdot\hat{q}\,\vec{\epsilon}_{d}\cdot\hat{q}
-\vec{\sigma}\cdot\vec{\epsilon}_{d}\right)
\,\varphi_{D}(q)\right]\mbox{\large $\eta$}_{pc}\:.
\end{eqnarray}

Here \mbox{\large $\eta$}$_n$ is the neutron spinor and
\mbox{\large $\eta$}$_{pc}$ the charge-conjugate proton spinor, and
$\varphi_{S}(q)$ and $\varphi_{D}(q)$ the deuteron S- and D-state components
in momentum space. When the deuteron is moving with energy $E_d$, the
longitudinal part of the argument of the wave function is changed to
\begin{equation}
\label{LorF}
(\vec{q}_{\perp},q_{\parallel})\longrightarrow
(\vec{q}_{\perp},q_{\parallel}/\gamma)\:,
\end{equation}
where $\gamma = E_d/m_d$.
\noindent
\subsection{The $^4\!H\!e$ wave function}

In the reaction model of fig.~1, we require the projection of the
$^4\!H\!e$ wave function onto a bound state of a neutron and
a $^3\!H\!e\,(h)$.
In terms of the neutron and charge-conjugate helion spinors, this component of
the final $^4\!H\!e$ wave function in momentum space is
\begin{equation}
{\cal M}(nh_{c}\to {\alpha}_{f})
=\half \sqrt{N_{\alpha}}\,\mbox{\large $\zeta$}_{hc}^{\dagger}\,
\psi_{\alpha}^*(k)\,\mbox{\large $\eta$}_{n}\:,
\end{equation}
where $\vec{k}$ is the relative neutron-helion momentum inside the
$^4\!H\!e$.

If we take a completely symmetric Gaussian four-particle wave function
for the $^4\!H\!e$ nucleus, then the relative wave function
between the neutron and helion is also Gaussian,
\begin{equation}
\psi_{\alpha}(\vec{k})=\left(\frac{\beta^2}{\pi}\right)^{\!3/4}\,
\exp(-\half\beta^2k^2)\:.
\label{nh-pair-alpha-wf}
\end{equation}
The parameter $\beta$ may be fixed by the point rms radius of $^4\!H\!e$,
\begin{equation}
<r^2>_{pt}= (1.46)^2=\mbox{${\textstyle \frac{27}{32}}$}\beta^2\:.
\end{equation}
This gives $\beta=1.59\,$fm.

In this simple model, the total number of $p\,^{3\!}H$ and
$n\,^{3\!}H\!e$ pairs in the $\alpha$-particle is $N_{\alpha}=4$.

\subsection{Pion production $pd\to\pi^0\,{^3\!H\!e}$}

There are only two independent $pd\to\pi^0\,{^3\!H\!e}$ amplitudes in the
forward/backward direction \cite{GW3}.
\begin{eqnarray}
{\cal M}(p_{c}d\to\pi^{0}h_{c})= \mbox{\large $\eta$}_{pc}^{\dagger}
\left[{\cal A}\,(\vec{\epsilon}_{d}\cdot\hat{p}_{d})
-i\,{\cal B}\,\vec{\sigma}\cdot(\vec{\epsilon}_{d}\times\hat{p}_{d})
\right]\mbox{\large $\zeta$}_{hc}\:,  \label{pionabsM}
\end{eqnarray}
where $\hat{p}_{d}$ is the direction of the deuteron  c.m.\ momentum.

This decomposition is invariant in all systems connected by a boost along the
beam direction. ${\cal A}$ represents the amplitude for helicity $m=0$
deuterons and ${\cal B}$ that for $m=\pm 1$ such that the aligned
$pd\to\pi^0\,{^3\!H\!e}$ c.m.~cross sections are given by
\begin{eqnarray}
\nonumber
\left[\frac{d\sigma}{d\Omega}(pd\to\pi^0\,{^3\!H\!e})\right]_{cm}^{m}
&=& \frac{m_{p}m_{h}}{4\,(2\pi)^{2}s_{pd}}\,
\frac{p_{\pi}}{p_p}\,{\mid}{\cal A}{\mid}^{2}\ \ \ (m=0)\:,\\[1ex]
&=& \frac{m_{p}m_{h}}{4\,(2\pi)^{2}s_{pd}}\,
\frac{p_{\pi}}{p_p}\,{\mid}{\cal B}{\mid}^{2}\ \ \ (m=\pm 1)\:.
\label{pd3Hepi-cross-sect}
\end{eqnarray}
\subsection{$(\pi,\eta)$ transmutation; $\pi^0n\to\eta n$}

The unique threshold amplitude for $\pi^0n\to\eta n$ is
\begin{equation}
{\cal M}(\pi^{0}n\to n'\eta)=
-\frac{1}{\sqrt{2}}\,
\mbox{\large $\eta$}_{n'}^{\dagger}{\cal G}(s_{\eta p})\,
\mbox{\large $\eta$}_{n}\:,
\end{equation}
where ${\displaystyle s_{\eta p}}$ is the square of the $\eta$-proton
c.m.~energy. The amplitude for the more measurable $\pi^-p\to\eta n$
reaction is a factor of $-\sqrt{2}$ larger, and the associated
unpolarised production cross section is given by
\begin{equation}
\left[\frac{d\sigma}{d\Omega}(\pi^-p\to\eta n)\right]_{cm}
= \frac{m_{p}^2}{4\,(2\pi)^{2}s_{\eta n}}\,
\frac{p_p}{p_{\pi}}\,{\mid}{\cal G}(s_{\eta n})\,{\mid}^{2}\:.
 \label{piNtoetaN-cross}
\end{equation}
\newpage
\noindent
\section{Linearisation of the propagator and}\vspace{-5mm}

\noindent
{\Large{\bf \ \ \ \ \ evaluation of the matrix element}}\\[2ex]
Just as for the $pd\rightarrow \,^3\!H\!e\,\eta$ reaction discussed in I,
our two-step mechanism for $dd\rightarrow \,^4\!H\!e\,\eta$ is such that the
intermediate state is almost on the energy shell, corresponding to a nearly
vanishing denominator in eq.(\ref{basM}), even for small Fermi momenta
$\vec{q}$ and $\vec{k}$.  A good approximation for the integral may then be
obtained by expanding this denominator around the point $\vec{q}=0$,
$\vec{k}=0$, retaining only linear terms.
All dependence of the $pd\to\pi^0\,{^3\!H\!e}$ and $\pi^0n\to\eta n$
amplitudes on the Fermi momenta are neglected when evaluating the nuclear
matrix element $\cal M_{K}$.

To first order in the Fermi momenta, the energy difference
\begin{equation}
\Delta E=E_{0}-E_{\rm int}=
\Delta E_{0}+\vec{k}\cdot\vec{W}+\vec{q}\cdot\vec{V}\: .
\end{equation}
Here
\begin{equation}
\Delta E_{0}=E_{\eta}(\vec{p}_{\eta})+E_{\alpha}(-\vec{p}_{\eta})
            -E_h(-\mbox{${\textstyle \frac{3}{4}}$}\vec{p}_{\eta})
            -E_{n}(\mbox{${\textstyle \frac{1}{2}}$}\vec{p}_d)
            -E_{\pi}(\mbox{${\textstyle \frac{3}{4}}$}\vec{p}_{\eta}
            -\mbox{${\textstyle \frac{1}{2}}$}\vec{p}_d)
\end{equation}
and the relativistic relative velocity vectors $\vec{V}$ and $\vec{W}$,
\begin{eqnarray}
\nonumber
\vec{V} &=& \vec{v}_{\pi}(\mbox{${\textstyle \frac{3}{4}}$}
\vec{p}_{\eta}-\mbox{${\textstyle \frac{1}{2}}$}\vec{p}_d)
- \vec{v}_{n}(\mbox{${\textstyle \frac{1}{2}}$}\vec{p}_d)\\[1ex]
&=&\frac{3}{4}\frac{1}{E_{\pi}(\frac{3}{4}\vec{p}_{\eta}-
\frac{1}{2}\vec{p}_d)}
\,\vec{p}_{\eta}-\frac{1}{2}\left[ \frac{1}{E_{\pi}
(\frac{3}{4}\vec{p}_{\eta}-
\frac{1}{2}\vec{p}_d)}+\frac{1}{E_{n}(\frac{1}{2}
\vec{p}_d)}\right]\vec{p}_d\:,
\nonumber \\[1ex] \nonumber
\vec{W}&=&-\vec{v}_{\pi}(\mbox{${\textstyle \frac{3}{4}}$}
\vec{p}_{\eta}-\mbox{${\textstyle \frac{1}{2}}$}\vec{p}_d)
+ \vec{v}_{h}(-\mbox{${\textstyle \frac{3}{4}}$}\vec{p}_{\eta})\\[1ex]
&=&-\frac{3}{4}\left[ \frac{1}{E_{\pi}(\frac{3}4\vec{p}_{\eta}-
\frac{1}{2}\vec{p}_d)}+\frac{1}{E_h(-\frac{3}{4}\vec{p}_{\eta})}\right]
\vec{p}_{\eta}+\frac{1}{2} \frac{1}{E_{\pi}(\frac{3}{4}\vec{p}_{\eta}-
\frac{1}{2}\vec{p}_d)} \,\vec{p}_d\:,
\label{VandW}
\end{eqnarray}
depend only upon external kinematic variables.

After replacing $m_n/E_n(\half\vec{p}_d)$ by $m_d/E_d(\vec{p}_d)$,
the linearisation simplifies the matrix element $\cal M$ of
eq.(\ref{basM}) to
\begin{eqnarray}
\nonumber
{\cal  M } = \frac{1}{(2 \pi)^3} \frac{1}{2E_{\pi}
(\frac{3}4\vec{p}_{\eta}-\frac{1}{2}\vec{p}_d)}\,
\frac{m_d}{E_{d}(\mbox{${\textstyle \frac{1}{2}}$}\vec{p}_d)}\,
\frac{m_h}{E_h(-\mbox{${\textstyle \frac{3}{4}}$}\vec{p}_{\eta})}\,N_T\,
\sqrt{4m_{\alpha}m_d}\,\\
\times\int  d^{\,3}k
\int d^{\,3}q\:\frac{i}{ \Delta E_{0}+\vec{k}\cdot\vec{W}+
\vec{q}\cdot\vec{V}+i\epsilon}\:{\cal M}_{K}\:. \label{appbasM}
\end{eqnarray}

Close to threshold the internal helion is almost at rest but relativistic
effects are important for the incoming deuteron. Taking them into account
leads to
\begin{equation}
{\cal  M } = \frac{1}{(2 \pi)^3} \frac{1}{2E_{\pi}
(\frac{3}4\vec{p}_{\eta}-\frac{1}{2}\vec{p}_d)}\,N_T\,\sqrt{4m_{\alpha}m_d}\,
\int  d^{\,3}k
\int d^{\,3}q\:\frac{i}{ \Delta E_{0}+\vec{k}\cdot\vec{W}+
\vec{q}\cdot\vec{V}^{\,'}+i\epsilon}\:{\cal M}_{K}\:, \label{appbasMp}
\end{equation}
where
\begin{eqnarray}
\nonumber
\vec{V}_{\perp}^{\,'} &=&\vec{V}_{\perp}\:,\\
V_{\parallel}^{\,'} &=& \gamma V_{\parallel}\:.
\end{eqnarray}

For production in the forward direction, the vectors $\vec{V}$ and $\vec{W}$
are parallel to the beam direction in the overall c.m.~frame,
\begin{equation}
\hat{V}^{\,'} = -\hat{p}_{d}\:,\ \  \hat{W} = \hat{p}_{d}\:.
\label{coll-kin}
\end{equation}
After neglecting the Fermi momentum, the $p d\to\pi^{0}\,^3\!H\!e$ amplitude
should be taken at $\theta_{p\pi}^{cm}=180^0$.

To evaluate the matrix element $\cal M$ of eq.(\ref{appbasMp}), we first
rewrite the denominator as an integral over a parameter $t$,
\begin{equation}
{\cal  M } =  \frac{1}{(2 \pi)^3}
\frac{1}{2E_{\pi}(\frac{3}4\vec{p}_{\eta}-\frac{1}{2}\vec{p}_d)} \,N_T\,
\sqrt{4m_{\alpha}m_d}\,
\int_{0}^{\infty} dt
\int  d^{\,3}k  \int d^{\,3}q \,e^{it (\Delta E+ i \epsilon)}
\,{\cal M}_{K}\:, \label{finalM}
\end{equation}
which allows the integral over the Fermi momenta in the wave functions to be
evaluated analytically.

The spin-algebra can be handled by the same tensor techniques used in I.
For collinear kinematics, (\ref{coll-kin}),
the matrix element ${\cal M}$ yields an expression for
the $F$ amplitude of eq.(\ref{uppdelM}) in terms of a single complex
form factor
\begin{equation}
\label{FormF}
S(\vec{W},\vec{V}^{'})= (2\pi)^3\sqrt{N_{\alpha}}\,
\int_{0}^{\infty} dt\,e^{it\Delta E_{0}}
\psi_{\alpha}^{*}(-t\vec{W})\left(\varphi_S(t\vec{V}^{'})+\frac{1}{\sqrt{2}}
\varphi_D(t\vec{V}^{'})\right)\:.
\end{equation}
This involves an integral over {\it configuration-space} $^4\!H\!e$ and $S$-
and
$D$-state deuteron wave functions.

The $F$ amplitude may be written in terms of this form factor as
\begin{equation}
F= - \frac{N_T\,{\cal G}\,{\cal B}\,S}
{4E_{\pi}(\half\vec{p}_d)}\:\cdot
\label{final-F}
\end{equation}
This is the  expected  structure since  both the  ${\cal B}$ and  $F$
amplitudes
correspond to incident deuterons with helicity $m=\pm 1$.

At threshold, we make use of eqs.(\ref{piNtoetaN-cross}) and
(\ref{pd3Hepi-cross-sect}) to replace the squares of the amplitudes ${\cal G}$
and ${\cal B}$ by the corresponding (polarised) cross sections. From
eqs.(\ref{final-F}) and (\ref{dd_cs}) it follows that
\begin{eqnarray}
\nonumber
\left[\frac{p_d}{p_{\eta}}\:\frac{d\sigma}{d\Omega}
(dd\rightarrow\,^4\!H\!e\,\eta)\right]_{cm} =
\frac{64}{3}\,\frac{1}{(2\pi)^{4}\,E_{\pi}^{2}(\frac{1}{2}\vec{p}_d)}\,Y_{M}\,
|S(\vec{W},\vec{V}^{\,'})|^{2}\times     \\ \nonumber  \\
\left[\frac{p_{d}}{p_{\pi}}\frac{d\sigma}{d\Omega}
(dp\to\,^3\!H\!e\,\pi^{0})\right]^{m=\pm1}_{cm} \:
\left[\frac{p_{\pi}}{p_{\eta}}\frac{d\sigma}{d\Omega}(\pi^{-}p\to\eta n)
\right]_{cm}\:,
\label{appr-f-dd}
\end{eqnarray}
where the value $N_T=12$ has been substituted.

The mass factor $Y_{M}$ is defined as
\begin{equation}
\label{mass_y}
Y_{M}=\frac{(m_{p}+m_{\eta})^2}{m_{p}^2}\,
\frac{(m_{\alpha}+m_{\eta})^{2}+\frac{1}{8}m_{\alpha}^2}
{(m_{\alpha}+m_{\eta})^{2}}\,\cdot
\end{equation}

The two factors of the expression for $Y_{M}$ are related to the c.m.~energies
of the participating $\pi^{+}n\to\eta p$ and $dp\to\,^3\!H\!e\,\pi^{0}$
reactions, and originate from terms in the
cross section formulas (\ref{piNtoetaN-cross}) and (\ref{pd3Hepi-cross-sect}).
It will be noted that the approximate threshold formula (\ref{appr-f-dd}) is
similar in structure to the corresponding one for the near-threshold
$pd\to\,^3\!H\!e\,\eta$ reaction derived in I.

In eq.(\ref{appr-f-dd}) the value of the $dp\to\,^3\!H\!e\,\pi^{0}$
cross section should be taken at $\theta_{p\pi}=180^{o}$, for
deuteron helicity $m=\pm 1$. Such cross sections have been well measured in our
energy domain \cite{Kerb}.

\newpage
\section{Final state interaction}

Though the experimental amplitude for $pd\rightarrow\,^3\!H\!e\,\eta$ varies
very rapidly near threshold \cite{Garcon}, the predictions of our two-step
model
are much smoother since the form factors hardly change for energy steps of a
few MeV. This sharp dependence was ascribed to a strong final state
interaction \cite{CW1}, and in I we developed a model for the S-wave
enhancement factor $\Omega$ based on an effective wave technique. Generalising
this to the case of four nucleons, it is straightforward to see that
the effective wave $\Psi_{i}(\vec{x_i})$ incident on a nucleon at
$\vec{x}_{i}$ is given by
\begin{equation}
\Psi_{i}(\vec{x}_{i})=e^{i\vec{k}\cdot\vec{x}_{i}}\:\left.
\left[1+ f\,\frac{e^{ikl}}{l} \,\left(\sum_{j\neq i}e^{i\vec{k}
\cdot(\vec{x}_{j}-
\vec{x}_{i})} -2\right) \right]\right/\! D \:,
\label{eqeffwave}
\end{equation}
for which the denominator is
\begin{equation}
D=\left(1+ f \,\frac{e^{ikl}}{l}\,\right)
\left(1-3 f \,\frac{e^{ikl}}{l}\,\right).
\end{equation}

In the above, $f$ is the $\eta N$ scattering amplitude and $k$ the momentum
of the $\eta$ scattering off the nucleons in the $^{4}\!H\!e$-nucleus, which
is assumed to be represented by four scattering centres placed at
the vertices of a tetrahedron with side length $l$.

The point $^4\!H\!e$ charge form factor in the rigid tetrahedron model is
\begin{equation}
F_{\rm ch}(q) = j_0\left(ql\sqrt{\frac{3}{8}}\right)\:.
\end{equation}
To obtain consistency between the $^4\!H\!e$ r.m.s.~radius and the form factor
minimum \cite{Frosch}, the Bessel function may be smeared over the
interparticle separation with a weight function
\begin{equation}
G(l) = N\,(l-l_{\mbox{\rm min}})\,\theta(l-l_{\mbox{\rm min}})\,
\exp(-\lambda^2(l-A)^2)\:,
\label{smearf}
\end{equation}
with parameter values
\begin{equation}
l_{\mbox{\rm min}} =0.50\,\mbox{\rm fm}\:,
\ \ \ A=0.82\,\mbox{\rm fm}\:,\ \ \ \
\lambda = 0.543\,\mbox{\rm fm}^{-1}\:.
\end{equation}
The subsidiary maximum in the form factor is then underpredicted by about a
factor of 1.7, but this feature is sensitive to meson exchange-current effects.

The $S$-wave enhancement factor becomes
\begin{equation}
\Omega=\left<\frac{1}{\mbox{\rm D}}\left [ 1+ f \,\frac{e^{ikl}}{l} \,
\left(\sum_{i>1}e^{i\vec{k}\cdot(\vec{x}_{i}-\vec{x}_{1})}
-2\right) \right]
\right>\:.
\label{eq:enh}
\end{equation}

Averaging over the orientation of the tetrahedron projects out the $S$-wave, to
leave an expectation value with respect to the lengths of its sides
\begin{equation}
\Omega = \int_{0}^{\infty}dl\,G(l)\left[\frac{1}{D}
\left(1+ f\,\frac{e^{ikl}}{l}\,\{3j_{0}(kl)-2\} \right)\right]\:.
\label{EF}
\end{equation}

Following the prescription used in I, for finite nucleon masses we
take the effective amplitude
\begin{equation}
f = \left(\frac{1+m_{\eta}/m_{N}}{1+m_{\eta}/m_{\alpha}}\right)
\,f_{\eta  N}\:.
\end{equation}
$f_{\eta N}$ is here the $\eta$-nucleon c.m.~scattering amplitude, evaluated
at a c.m.~momentum,
\begin{equation}
k_{\eta N}=\left(\frac{1+m_{\eta}/m_{\alpha}}
{1+m_{\eta}/m_{N}}\right)\:k\:,
\end{equation}
where $k$ is the c.m.~momentum in the $^4\!H\!e\,\eta$ system

In fig.~2 we show the variation of $\mid \Omega\mid^2$ with $p_{\eta}$ in the
illustrative case where the $\eta$-nucleon scattering amplitude is taken to be
constant at the scattering length value $a=(0.476+0.279i)\,$fm used in I.  At
threshold the attractive $\eta$-nucleon interaction enhances the predictions by
a factor $\mid \Omega\mid^2 = 1.9$. However the rapid fall-off from
enhancement to suppression around 25~MeV/c is very striking.
\newpage
\section{The $n\,^3\!H\!e\rightarrow \,^4\!H\!e\,\eta$ reaction}

If the near-threshold energy dependence is characteristic of an
$\eta\,^{4\!}H\!e$ final state interaction, then the effect should be largely
independent of the particular entrance channel. An alternative means of
reaching
this final state is through studying the
$n\,^3\!H\!e\rightarrow \,^4\!H\!e\,\eta$ reaction or the equivalent one with a
proton beam and tritium target.

Just as for the $dd$ case, there is only one amplitude at threshold or in the
forward direction, and it may be written as
\begin{equation}
{\cal M} = -\frac{1}{(2\pi)^3}\,\frac{1}{2\sqrt{6}}\,
\sqrt{\frac{m_{\alpha}}{m_d}}\,H\,
\mbox{\large $\zeta$}_{hc}^{\dagger}\,\vec{\sigma}\cdot\hat{p}_n\,
\mbox{\large $\eta$}_{n}\:.
\end{equation}

In terms of the amplitude ${\cal M}$, the differential cross section  becomes
\begin{equation}
\left[\frac{d\sigma}{d\Omega}(n\,^3\!H\!e\rightarrow
\,^4\!H\!e\,\eta)\right]_{cm}
= \frac{m_pm_h}{(2\pi)^{2}\,16s}\, \frac{p_{\eta}}{p_n}
\,\sum_{spins}{\mid}{\cal M}{\mid}^{2}\:.\label{nh_cs}
\end{equation}

The two-step model for the $n\,^3\!H\!e\rightarrow \,^4\!H\!e\,\eta$ reaction
shown in fig.~3, merely involves a rearrangement of the initial nucleons in
fig.~1. Thus ${\cal M}_{K}$ contains the same type of matrix elements as
eq.(\ref{nuclMk}) with the exception of the initial wave function,
\begin{equation}
{\cal M}_{\rm K}=\sum_{\mbox{\scriptsize int}}\,
            {\cal M}(h_{ic}\to p_{c}'d)\,
            {\cal M}(\pi^{0}p_{c}'\to p_{c}\eta_{f})\,
            {\cal M}(p_c t\to {\alpha}_{f})\,
            {\cal M}(n_id\to\pi^{0}t)\: .
\label{nuclMnh}
\end{equation}

After carrying out the linearising procedure as in \S 3, it is straightforward
to show that the two-step contribution to the amplitude $H$ is
\begin{equation}
H= \frac{3{\cal GA}}{2E_{\pi}(\frac{1}{3}\vec{p}_n)}
\left(S_S(\vec{W},\vec{V}')-\sqrt{2}S_D(\vec{W},\vec{V}')\,
\right)
+\frac{3{\cal GB}}{E_{\pi}(\frac{1}{3}\vec{p}_n)}\,
\left(S_S(\vec{W},\vec{V}')+S_D(\vec{W},\vec{V}')/\sqrt{2}\right)
\:,
\label{final-Fnh}
\end{equation}
where an extra factor of three arises from including also the analogous
charged pion contribution in fig.~3.

The form factors are defined by
\begin{equation}
S_{S,D}(\vec{W},\vec{V}^{'})=(2\pi)^3\,\sqrt{N_{\alpha}N_d}\,
 \int_{0}^{\infty} dt\,e^{it\Delta E_{0}}
\psi_{\alpha}^{*}(-t\vec{W})\phi_{S,D}(t\vec{V}^{'})\:,
\end{equation}
where $\phi_{S}(\vec{r})$ and $\phi_{D}(\vec{r})$ are the helion S- and
D-state wave functions of the $pd$ partition, a parametrisation of which is
given in ref.\cite{GW1}, and the number of such partitions $N_d\approx 1.5$.

Here the bare intermediate state energy difference is given by
\begin{equation}
\Delta E_{0}=E_{\eta}(\vec{p}_{\eta})+E_{\alpha}(-\vec{p}_{\eta})
            -E_t(-\mbox{${\textstyle \frac{3}{4}}$}\vec{p}_{\eta})
            -E_{p}(\mbox{${\textstyle -\frac{1}{3}}$}\vec{p}_n)
            -E_{\pi}(\mbox{${\textstyle \frac{3}{4}}$}\vec{p}_{\eta}
            +\mbox{${\textstyle \frac{1}{3}}$}\vec{p}_n)
\end{equation}
and the relativistic velocity vectors $\vec{W}$ and $\vec{V}^{'}$
are defined as
\begin{eqnarray}
\nonumber
\vec{V}^{'} &=& \frac{E_{\rm h}(-\vec{p}_n)}{m_{\rm h}}\left[
\vec{v}_{\pi}(\mbox{${\textstyle \frac{3}{4}}$}
\vec{p}_{\eta}+\mbox{${\textstyle \frac{1}{3}}$}\vec{p}_n)
- \vec{v}_{p}(-\mbox{${\textstyle \frac{1}{3}}$}\vec{p}_n)\right] \:,
\nonumber \\[1ex]
\vec{W}&=&\vec{v}_{t}(-\mbox{${\textstyle \frac{3}{4}}$}
\vec{p}_{\eta})
-\vec{v}_{\pi}(\mbox{${\textstyle \frac{3}{4}}$}
\vec{p}_{\eta}+\mbox{${\textstyle \frac{1}{3}}$}\vec{p}_n)
\:.
\end{eqnarray}

A crucial difference with the $dd\rightarrow \,^4\!H\!e\,\eta$
calculation is that the input $pd\to\,^3\!H\!e\,\pi^0$ amplitudes are
required in the \underline{forward} direction but at a rather higher incident
energy. The experimental data \cite{Kerb} indicate that
\begin{equation}
\mid {\cal A}(\theta_{p\pi}=0)\mid^2\ >>\
\mid {\cal B}(\theta_{p\pi}=0)\mid^2\:.
\end{equation}

If we only keep the dominant ${\cal A}$ amplitude, then we can derive a linear
relation between the $n\,^{3\!}H\!e\rightarrow \,^4\!H\!e\,\eta$ and the
{\it unpolarised} input cross sections
\begin{eqnarray}
\nonumber
\left[\frac{p_n}{p_{\eta}}\:\frac{d\sigma}{d\Omega}
(n\,^{3\!}H\!e\rightarrow\,^4\!H\!e\,\eta)\right]_{cm} =
\frac{81}{128}\,\frac{1}{(2\pi)^{4}\,E_{\pi}^{2}(\frac{1}{3}\vec{p}_n)}\,
Y_{M}'\,
|S_S(\vec{W},\vec{V}^{\,'}) -\sqrt{2}\,S_D(\vec{W},\vec{V}^{\,'})
|^{2}\times     \\ \nonumber  \\
\left[\frac{p_{d}}{p_{\pi}}\frac{d\sigma}{d\Omega}
(pd\to\,^3\!H\!e\,\pi^{0})\right]_{cm} \:
\left[\frac{p_{\pi}}{p_{\eta}}\frac{d\sigma}{d\Omega}(\pi^{-}p\to\eta n)
\right]_{cm}
\label{n3he}
\end{eqnarray}

The mass factor $Y_{M}'$ is similar to that given in eq.(\ref{mass_y}),
\begin{equation}
Y_{M}' = \frac{(m_{p}+m_{\eta})^2}{m_{p}^2}\:
\frac{m_{\alpha}^{2}+\frac{32}{27}m_{\eta}(2m_{\alpha}+m_{\eta})}
{(m_{\alpha}+m_{\eta})^{2}}\,\cdot
\end{equation}

It is of course to be expected that the final state enhancement factor
$\mid\Omega\mid^2$ should be the same here as for the
$dd\rightarrow \,^4\!H\!e\,\eta$ reaction.
\newpage
\section{Predictions and Conclusions}

The Breit-Wigner fit to the $\pi^-p\to\eta n$ threshold amplitude given in
\cite{FW1} yields
\begin{equation}
\left[\frac{p_{\pi}}{p_{\eta}}\frac{d\sigma}{d\Omega}(\pi^{+}n\to\eta p)
\right]_{cm} = (680\pm 60)\,\mu b/sr\: ,
\end{equation}
where the error bar is taken from the direct measurement of
Binnie {\it et al.} \cite{Binnie}.

At the $dd\rightarrow \,^4\!H\!e\,\eta$ reaction threshold, the laboratory
deuteron kinetic energy is 1.12 GeV, and for this energy the helicity $m=1$
differential cross section of the $pd\to\,^3\!H\!e\,\pi^0$ reaction at
$\theta_{pd}^{cm}=180^0$ is measured to be \cite{Kerb},
\begin{equation}
\left[\frac{p_{d}}{p_{\pi}}\frac{d\sigma}{d\Omega}
(pd\to\,^3\!H\!e\,\pi^{0})\right]^{M=1}_{cm} =
(1.0\pm 0.1)\,\mu{\rm b}/{\rm sr} \:.
\end{equation}

Using the Gaussian parametrisation (\ref{nh-pair-alpha-wf}) of the
$^4\!H\!e:n^3\!H\!e$ cluster wave function, together with the Paris deuteron
wave function \cite{Paris}, the magnitude of the transition form
factor of eq.(\ref{FormF}) is found to be
\begin{equation}
|S| = 11.2\,\mbox{\rm fm}^{-2}
\end{equation}
at the threshold for $\eta$-production.

Before introducing final state interactions into our model, eq.(\ref{dd_cs})
predicts a threshold amplitude squared of
\begin{equation}
\left|f(dd\rightarrow \,^4\!H\!e\,\eta)\right|^2 = \left[\frac{p_d}{p_{\eta}}\,
\frac{d\sigma}{d\Omega}(dd\rightarrow \,^4\!H\!e\,\eta)\right]_{cm}
= (44\pm 8)\,\mbox{\rm nb/sr}\, ,
\end{equation}
to be compared with the experimental value \cite{Roudot} of about
$30$~nb/sr.

The energy dependence of the two-step form factor and the input
$dp\to\,^3\!H\!e\,\pi^{0}$ cross section are both negligible for small
$p_{\eta}$. The only significant variations arise from the final state
interactions and, to a lesser extent, the $\pi^{-}p\to\eta n$ amplitude
squared.

After including the final state interaction effects, using the formalism of
\S 4, the predictions shown in fig.~4, which for ease of comparison have been
multiplied by a scale factor of $0.75$, seem to have too sharp a momentum
dependence as
compared to the SPESIV data \cite{Roudot}. Though these data were taken over a
very small range in $p_{\eta}$, and the lowest point is subject to large
corrections due to energy losses in the target and similar effects, the much
smaller slope as compared to $pd\rightarrow \,^3\!H\!e\,\eta$ is confirmed in
the SPESIII experiment over a much wider domain in $p_{\eta}$ \cite{Amina}.
It should however be remarked that the FSI factor $|\Omega|^2$ depends
sensitively upon the $\eta$-nucleon scattering length and it has recently been
suggested \cite{Svarc} that it might be even larger than that which we have
taken here.

In the case of the $n\,^3\!H\!e\rightarrow \,^4\!H\!e\,\eta$ reaction, the
neutron laboratory kinetic energy at threshold is $0.75$~GeV so that the
$dp\to\,^3\!H\!e\,\pi^{0}$ input is required at laboratory kinetic energy of
$T_d = 1.50$~GeV for $\theta_{p\pi}^{cm}=0^0$.
Using the data of ref.\cite{Kerb}, we find
\begin{equation}
\left[\frac{p_{d}}{p_{\pi}}\frac{d\sigma}{d\Omega}
(pd\to\,^3\!H\!e\,\pi^{0})\right]^{unpol}_{cm} =
(2.8\pm 0.3)\,\mu{\rm b}/{\rm sr} \:.
\end{equation}

Taking the Germond-Wilkin parametrisation of the $^3\!H\!e:dp$ cluster wave
function \cite{GW1}, we find the magnitude of the form factor to be
\begin{equation}
\left|S_S -\sqrt{2}\,S_D\right| = 17.3\ \mbox{\rm fm}^2\:,
\end{equation}
so that the predicted amplitude squared becomes
\begin{equation}
\left|f(n\,^3\!H\!e\rightarrow \,^4\!H\!e\,\eta)\right|^2
= \left[\frac{p_d}{p_{\eta}}\,
\frac{d\sigma}{d\Omega}(n\,^3\!H\!e
\rightarrow \,^4\!H\!e\,\eta)\right]_{cm}
= (21\pm 4)\,\mbox{\rm nb/sr}\, ,
\end{equation}
which is about a factor of two smaller than the $dd$ prediction. One of the
major reasons for the difference between the two predictions is the much larger
numerical factor in eq.(\ref{appr-f-dd}) compared to (\ref{n3he}). In the
$dd$ case there are four times as
many graphs, all of which contribute coherently.

Unlike the $dd$ case, the two-step model for the
$n\,^3\!H\!e\rightarrow \,^4\!H\!e\,\eta$ reaction in fig.3 should have
contributions from intermediate spin-singlet $NN$ ($d^*$) states. From the
experience gathered in I, these could well increase the prediction by over a
factor of two, to make it comparable to the $dd$ result.
Nevertheless, it is much harder to investigate this reaction
experimentally due to the difficulties of working with a broad-band neutron
beam
or a radioactive tritium target. Attempts at such a measurement at LAMPF were
aborted due to problems with the tritium target \cite{Ben}.

There are however few other possibilities.
The low energy $\eta\,^3\!H\!e$ system could be investigated through pion
production or even more cleanly through the photoproduction
$\gamma\,^3\!H\!e\to\eta\,^3\!H\!e$ since sufficient energy resolution can now
be obtained \cite{Bernd}. On the other hand photoproduction of the
$\eta\,^4\!H\!e$ system is strongly suppressed due to the dominance of
isovector
photon coupling to the $N^*(1535)$.

It should be noted that the near-threshold kinematics are not as `magic'
as they are for $pd\rightarrow \,^3\!H\!e\,\eta$, since in the case of
$dd\rightarrow \,^4\!H\!e\,\eta$ the zero-Fermi-momentum energy difference
$\Delta E_0 = -135$~MeV while for the
$n\,^{3}\!H\!e\rightarrow \,^3\!H\!e\,\eta$ case,
$\Delta E_0 = +140$~MeV. The imaginary parts of the form factors become rather
important, and the situation is quite similar to the
$pd\rightarrow \,^3\!H\!e\,(\omega,\eta')$ cases, which are successfully
reproduced by the two-step model \cite{FW2}.

In contrast to the $pd\rightarrow \,^3\!H\!e\,\eta$ estimate, our calculation
of
the   $dd\rightarrow    \,^4\!H\!e\,\eta$   amplitude is   slightly  higher
than
experiment.  Since   contributions from  all  constituents  in the  nuclear
wave
function have been  included, this reinforces the  suggestion made in I that
the
$pd$ discrepancy is  due to the  omission of the  intermediate $d^*$ effects.
Of
course  initial-state damping, which we  have not  considered, could well
reduce
somewhat the $^4\!H\!e\,\eta$ production rate.
\newpage
\noindent
{\large \bf Acknowledgements}\\

Leonardo Castillejo was a source of inspiration and advice for many years and
we
were saddened by his sudden death earlier this year.

One of the authors (CW) is indebted to the Royal Swedish Academy of Sciences
for continued support and the The Svedberg Laboratory for hospitality during
his stay in Uppsala. Discussions with N.~Willis, A.~Zghiche and Y.~Le Bornec
on the material of ref.\cite{Amina} are greatly appreciated.

\newpage
\vspace*{-2cm}
\begin{center}
{\Large{\bf Figure Captions}}\\[4ex]
\end{center}

\noindent
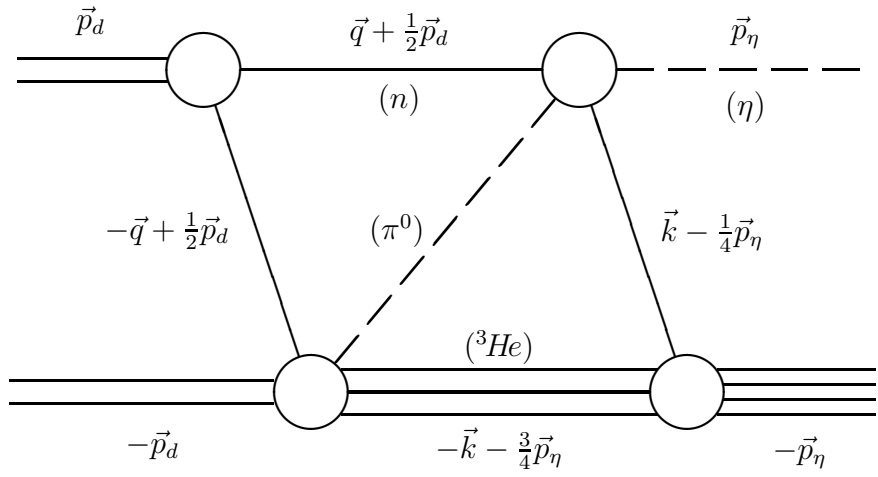
Figure~1 : Double-scattering diagram for
$dd\rightarrow \,^4\!H\!e\,\eta$
in terms of the constituent $pd\to\,^3\!H\!e\,\pi^0$
and $\pi^0 n\to \eta n$
amplitudes.
Single lines are nucleons, double deuterons, triple
$^3\!H\!e$ and quadruple
$^4\!H\!e$, the dashed
meson lines being labelled individually. The momentum labels
refer to the
overall c.m.~system. In total there are eight such diagrams corresponding to
the
interchanges $n,\,^3\!H\!e\to p,\,^3\!H$, $\pi^0 \to \pi^+$ and exchanging beam
and target deuterons.
\\[4ex]
Figure~2 : Enhancement factor $\mid\! \Omega\!\mid^2$ of
eq.(\protect\ref{EF})
evaluated in the constant $\eta$-N scattering length
approximation.\\[4ex]
Figure~3 : Double-scattering diagram for
$n\,^{3\!}H\!e\rightarrow \,^4\!H\!e\,\eta$
in terms of the constituent $pd\to\,^3\!H\!e\,\pi^0$
and $\pi^0 n\to \eta n$ amplitudes.\\[4ex]
Figure~4 : Spin-averaged amplitude squared
$\displaystyle |f|^2 = \left[\frac{p_d}{p_{\eta}}\,
\frac{d\sigma}{d\Omega}(dd\rightarrow \,^4\!H\!e\,\eta)\right]_{cm}$
for the $dd\rightarrow\,^4\!H\!e\,\eta$ reaction near threshold,
as a function
of the $\eta$ c.m.~momentum $p_{\eta}$. The theoretical curve of
eq.(\ref{dd_cs}), with FSI enhancement and multiplied by a constant factor of
$0.75$, is compared with
experimental data taken from ref.\cite{Roudot}.
\newpage
\begin{figure}[h]
\setlength{\unitlength}{1mm}
\begin{picture}(120,65)(-5,-10)
\thicklines
%
%
\put(100,0){\circle{10}}
\put(50,0){\circle{10}}
\put(85.71,42.86){\circle{10}}
\put(35.71,42.86){\circle{10}}
%
%
\put(98.42,4.74){\line(-1,3){11.12}}
\put(112.85,21.43){\makebox(0,0)[r]{$\vec{k}-\fmn{1}{4}\vec{p}_{\eta}$\ \ }}
\put(48.42,4.74){\line(-1,3){11.12}} \put(19.85,21.43){\makebox(0,0)[l]
          {\ \  $-\vec{q}+\fmn{1}{2}\vec{p}_d$}}
%
%
\put(80.71,42.86){\line(-1,0){40}}
\put(61.71,42.86){\makebox(0,10){$\vec{q}+\fmn{1}{2}\vec{p}_d$}}
\put(61.71,38.86){\makebox(0,0){$(n)$}}
\put(28.91,-2){\makebox(0,-10){$-\vec{p}_d$}}
%
%
\multiput(90.71,42.86)(6.8271,0){5}{\line(1,0){4.8271}}
\put(107.78,42.86){\makebox(0,10){$\vec{p}_{\eta}$}}
\put(107.78,42.86){\makebox(0,-10){($\eta$)}}
\multiput(53.20,3.84)(4.3706,5.2448){7}{\line(5,6){3.09}}
\put(67.85,21.43){\makebox(0,0)[r]{($\pi^{0}$)\ \ }}
%
%
\multiput(104,-3)(0,6){2}{\line(1,0){21}}
\multiput(104.9,-1)(0,2){2}{\line(1,0){20.1}}
%
%
\put(75,-2.5){\makebox(0,-10){$-\vec{k}-\fmn{3}{4}\vec{p}_{\eta}$}}
\put(75,11.0){\makebox(0,-10){($^{3\!}H\!e$)}}
%
%
\multiput(30.94,41.36)(0,3){2}{\line(-1,0){20}}
\put(20.71,44.36){\makebox(0,10){$\vec{p}_d$}}
\multiput(45.03,-1.5)(0,3){2}{\line(-1,0){35.17}}
\multiput(96.0,-3)(0,6){2}{\line(-1,0){42}}
\put(95,0){\line(-1,0){40}}
\put(115,-3){\makebox(0,-10){$-\vec{p}_{\eta}$}}

\end{picture}
\end{figure}
\vspace*{1cm}

\begin{center}
{\Large\sf Fig. 1}
\end{center}
\vspace{1cm}
\begin{figure}[h]
\setlength{\unitlength}{1mm}
\begin{picture}(120,65)(-5,-10)
\thicklines
%
%
\put(100,0){\circle{10}}
\put(50,0){\circle{10}}
\put(85.71,42.86){\circle{10}}
\put(35.71,42.86){\circle{10}}
%
%
\put(98.42,4.74){\line(-1,3){11.12}}
\put(112.85,21.43){\makebox(0,0)[r]{$\vec{k}-\fmn{1}{4}\vec{p}_{\eta}$\ \ }}
\put(50.0,5.0){\line(-1,3){11.34}}
\put(47.07,4.05){\line(-1,3){11.34}}

\put(19.85,21.43){\makebox(0,0)[l]
          {\ \  $-\vec{q}-\fmn{2}{3}\vec{p}_n$}}
%
%
\put(80.71,42.86){\line(-1,0){40}}
\put(61.71,42.86){\makebox(0,10){$\vec{q}-\fmn{1}{3}\vec{p}_n$}}
\put(61.71,38.86){\makebox(0,0){$(p)$}}
\put(28.91,-2){\makebox(0,-10){$\vec{p}_n$}}
%
%
\multiput(90.71,42.86)(6.8271,0){5}{\line(1,0){4.8271}}
\put(107.78,42.86){\makebox(0,10){$\vec{p}_{\eta}$}}
\put(107.78,42.86){\makebox(0,-10){($\eta$)}}
\multiput(53.20,3.84)(4.3706,5.2448){7}{\line(5,6){3.09}}
\put(67.85,21.43){\makebox(0,0)[r]{($\pi^{0}$)\ \ }}
%
%
\multiput(104,-3)(0,6){2}{\line(1,0){21}}
\multiput(104.9,-1)(0,2){2}{\line(1,0){20.1}}
%
%
\put(75,-2.5){\makebox(0,-10){$-\vec{k}-\fmn{3}{4}\vec{p}_{\eta}$}}
\put(75,11.0){\makebox(0,-10){($^{3\!}H$)}}
%
%
\put(30.71,42.86){\line(-1,0){19.77}}
\multiput(31.38,40.36)(0,5){2}{\line(-1,0){20.44}}
\put(20.71,44.36){\makebox(0,10){$-\vec{p}_n$}}
\put(45.0,0.0){\line(-1,0){35.14}}
\multiput(96.0,-3)(0,6){2}{\line(-1,0){42}}
\put(95,0){\line(-1,0){40}}
\put(115,-3){\makebox(0,-10){$-\vec{p}_{\eta}$}}
\end{picture}
\end{figure}
\vspace{1cm}

\begin{center}
{\Large\sf Fig. 3}
\end{center}
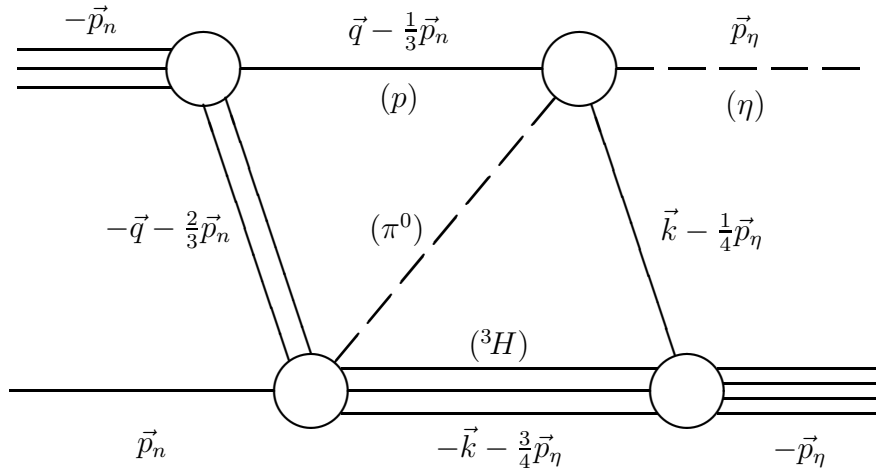
\end{document}